# An efficient multi-language Video Search Engine to facilitate the HADJ and the UMRA

**Mohamed HAMROUN**
LABRI
University of Bordeaux, France
mohamed.hamroun@u-bordeaux.fr

**Sonia LAJMI**
MIRACL
University of Sfax, Tunisia
Albaha University, Saudi Arabia
slajmi@bu.edu.sa

## Abstract

Videos clips  became the most important and prominent multimedia document to illustrate the rituals process of Hajj and Umrah. Therefore, it is necessary to develop a system to facilitate access to information related to  the duties, the pillars, the stages and the prayers.
In this paper present a new project accomplishing a search engine in a large video database enabling any pilgrims  to get the information that he care about as fast, accurate.
This project is based on two techniques: (a) the weighting method to determine the degree of affiliation of a video clip to a particular topic (b) organizing data using several layers.

## 1.    INTRODUCTION

When Muslims visit the two holy-Mosques to perform Hajj and Umrah, they needs information about rituals: the duties, the pillars, the stages and the prayers especially if it is here first time. In this context, Video clips are the most prominent media to illustrate the rituals. Therefore, it became necessary to develop an efficient video search engine to bring the image of the rituals for pilgrims.

On the other hand, thanks to advances in multimedia technology and lower storage costs, multimedia content archives have increased dramatically. With this exponential growth of content, problems emerge such as storage, access, search, navigation and extraction. Access to multimedia documents must be easy and fast. To do this, a process of searching and manipulating multimedia documents must be established.
In this paper we present a new approach to indexing and searching in a large video database. Specifically, we are interested by the informational videos related to the rituals of Hajj and Umra.
This paper is organized as following : after presenting the related works, we detail in the section 3 our proposed approach. The developed system is presented in the next section intituled multilingual search engine (English/Arabic).We detail after the test and experimentation. We present finally the conclusion and future works.

## 2.    RELATED WORKS

Research work on multimedia retrieval has been very varied and has been treated with different modalities (low levels and semantics).

Retrieval approaches based on low level characteristics remains limited despite the fact that they are numerous and complex [1]. These approaches are based on video segmenting (i.e, video became a set of keyframes), indexing and extracting low-level features such as shape, color, and textures of the keyframes. The semantic aspect such as the objects present in the keyframes are the most difficult to extract.

In the last decade, semantic retrieval became subject of several research works. For example, INFORMEDIA[2] approach uses a limited set of high-level concepts to filter the results of textual queries, which creates keyframe groups [3]. It exploits the results of automatic speech recognition to plot collections of keyframes at the associated geographic location on a map and combine this with other visualizations to give the user a contextual understanding of the outcome of a speech- a request. The method proposed in [4] is based on a semantic indexing process. This system uses a wide semantic lexicon, categories and "threads" to support interaction. It defines a space of visual similarity, a space of semantic similarity, a space of semantic "thread" and browsers to exploit these spaces. Let us also mention the VERGE approach [5], which supports the following functions: search by high-level visual concept and textual search. This tool combines indexing, analysis and recovery techniques in a variety of ways (textual, visual and conceptual).

As mentioned above, many works study the field of audiovisual data retrieval in a large database and offer tools based on various forms of research. The limit of these techniques often lies in the failure to take account of interactions with the user. Putting the user at the heart of the research process potentially offers more relevant results. In the literature, few studies are interested in this aspect.

The implementation of a semi-automatic approach, thus effectively putting the user at the center of the indexing and retrieval process, is therefore a real prospect of improving the existing methods.

## 3. PROPOSED APPROACH

Our approach aims to index and retrieve in a large database of a video documents. The Figure 1 illustrate the proposed approach of indexing and retrieval.

The process is composed by two stages (I) an indexation phase: this phase is realised in upstream : prepared in the database and updated occasionally when new videos are added to the system. (II) A search phase: this phase consist of indexing the query, comparing with the existing index extract from the last phase and return a list of audiovisual documents sorted by pertinence.

As illustrated in Figure 3, the process start by: (1) Indexing the query. Stop words are eliminated in the first. The weight of the rest of the terms is considered the same. (2) Make a comparison between the query terms and terms which are assigned to describe each concept among our database and returns concepts that have similar descriptors to the query. (3) In this step, the user operates to interactively choose the corresponding concepts. (4) Mapping between the query concepts and those of the whole collection is done using a vector model. This model incorporates a vector space representation which symbolizes documents or queries.

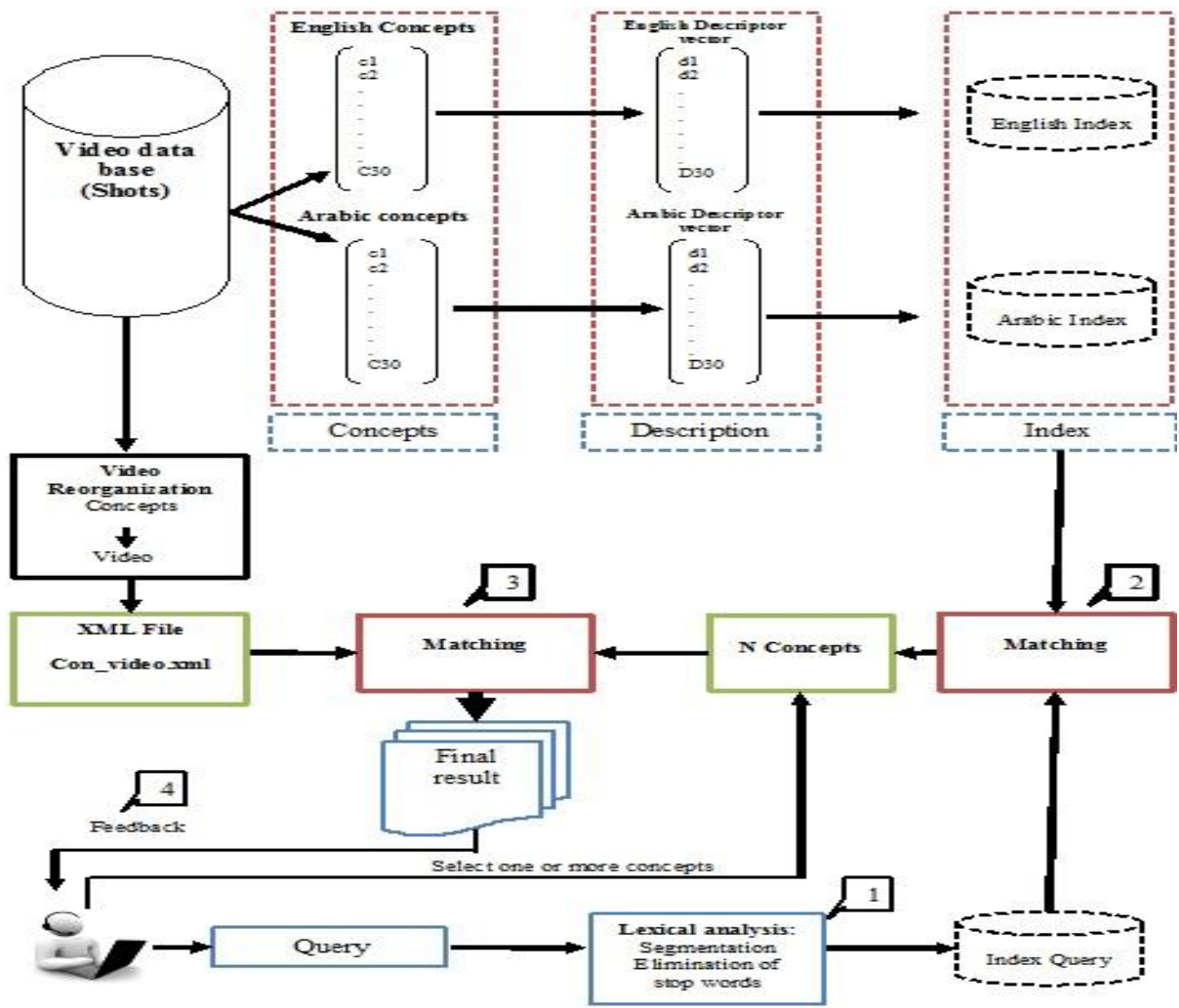

**Figure 1. Conceptual Architecture our indexing and retrieval system**

The advantage of this model resides in its simplicity. The cosine measure [6] is used to calculate the degree of relevance of each video document. This measure is the most successful and pertinent in the vector model.

In case of insatisfaction with the returned results, user can improve the result with the query expansion technique. In fact, when the user chose an item form the list of returned result, he can express explicitly the relevance of this item. Thus, the system recalculates and gives new results to the user.

### 3.1   ORGANIZATION OF THE DATABASE

In the indexation phase, videos are segmented in several keyframes. The segmentation is applied according to the model presented in [7]. When the keyframes are ready, an XML file is prepared for each set of keyframes. The XML file have the structure presented below (i.e, in Figure 2). To populate each XML document with concepts describing the audiovisual documents, we use a domain Ontology. The organization of the data into an XML file aim to organize the data and facilitate access to them. As described in the example in Figure 2, Each concept describing the audiovisual document has a Weight indicating the importance of this concept in description of this video. To calculate the weight we use TF-IDF measurement [6].

```xml
<?xml version="1.0" encoding="UTF-8" ?>
<concept>
- <videoFeatureExtractionFeatureResult fNum="1">
    <item seqNum="1" shotId="shot11176_10" />
    <item seqNum="2" shotId="shot11176_11" />
    <item seqNum="3" shotId="shot11176_12" />
    <item seqNum="4" shotId="shot11176_3" />
    <item seqNum="5" shotId="shot11176_4" />
    <item seqNum="6" shotId="shot11176_5" />
    <item seqNum="7" shotId="shot11176_6" />
    <item seqNum="8" shotId="shot11176_7" />
    <item seqNum="9" shotId="shot11176_8" />
    <item seqNum="10" shotId="shot11176_9" />
    <item seqNum="11" shotId="shot11249_10" />
    <item seqNum="12" shotId="shot11249_11" />
    <item seqNum="13" shotId="shot11249_12" />
```

**Figure 2. XML file describing videos by several concepts.**

### 3.2 STRUCTURING DATA

The data is structured according to three levels of abstraction: (i) the contextual level,(ii) the conceptual level and (iii) the data level. Therefore, our idea is to bring together all the data having common features and have a common concepts. For example, we assume that:

- The features of a video (V1) are: "طواف",
- The features of a video (V2) are: "عرفة",
- The features of a video (V3) are: "جمرات".

The transition to another level of abstraction helps us to organize the data and to accelerate access. Similarly, we grouped semantically most similar concepts in the same context for the extraction of concepts from context. We will study the same example to explain the transition to the third level:

- "طواف" belongs to"حج",
- "طواف" belongs to"عمرة",
- "عرفة" belongs to"حج",
- "جمرات" belongs to"حج".

General representation of our database is as follows:

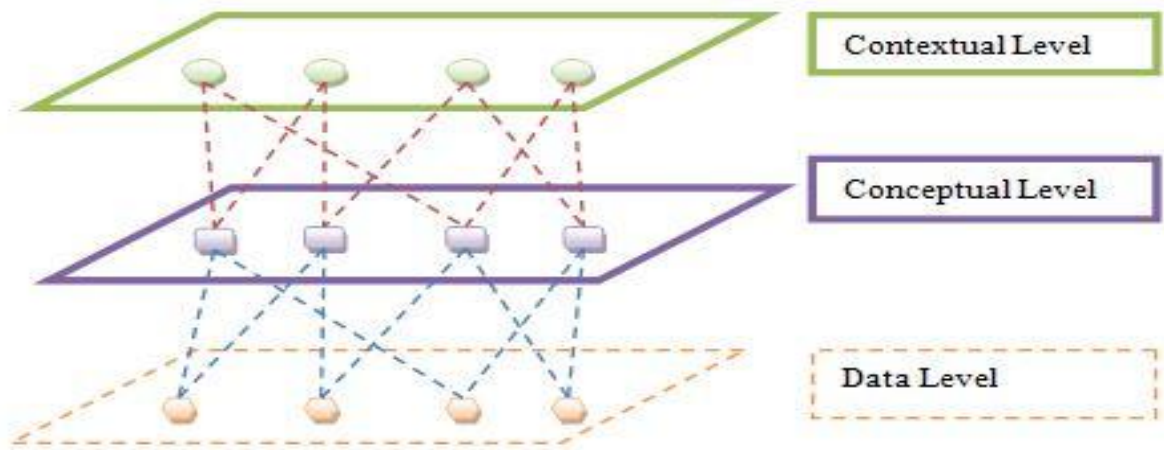

Figure 3. Database Structure

### 2.3 Query expansion

The most implicit way to retrieve videos is to express a textual query. Subsequently, the system analyzes the expressed query and translates it into concepts or features describing videos. In this context, the big problem is how we can translate a textual query into concepts?

To solve this problem, the query expansion technique is applied by employing a domain ontology to assist the user in formulating his queries. The Figure 4 represent a part of domain ontology proposed to the user to expand his query.

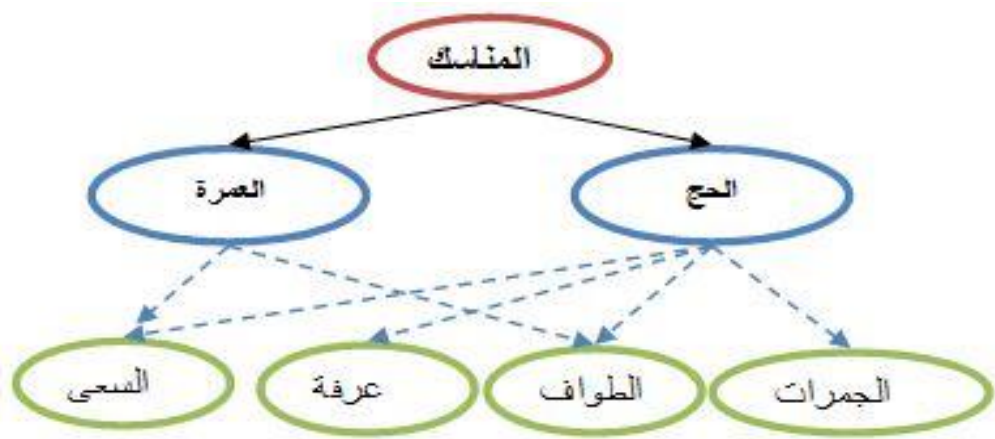

Figure 4. Part of concept of Ontology

### 4.    MULTILINGUAL SEARCH ENGINE (ENGLISH/ARABIC)

In this section we will detail search system (Arabic/English). Usually, the easiest way for  a user to express  his  request is a text that defines the  content of  videos which he want to see. The problem here is how to translate a textual query into concepts? To solve this problem we use the technique of query expansion by the call for ontology to help users to formulate their queries. Figure 4 explains the different stages of the research proposed system, based on text queries.

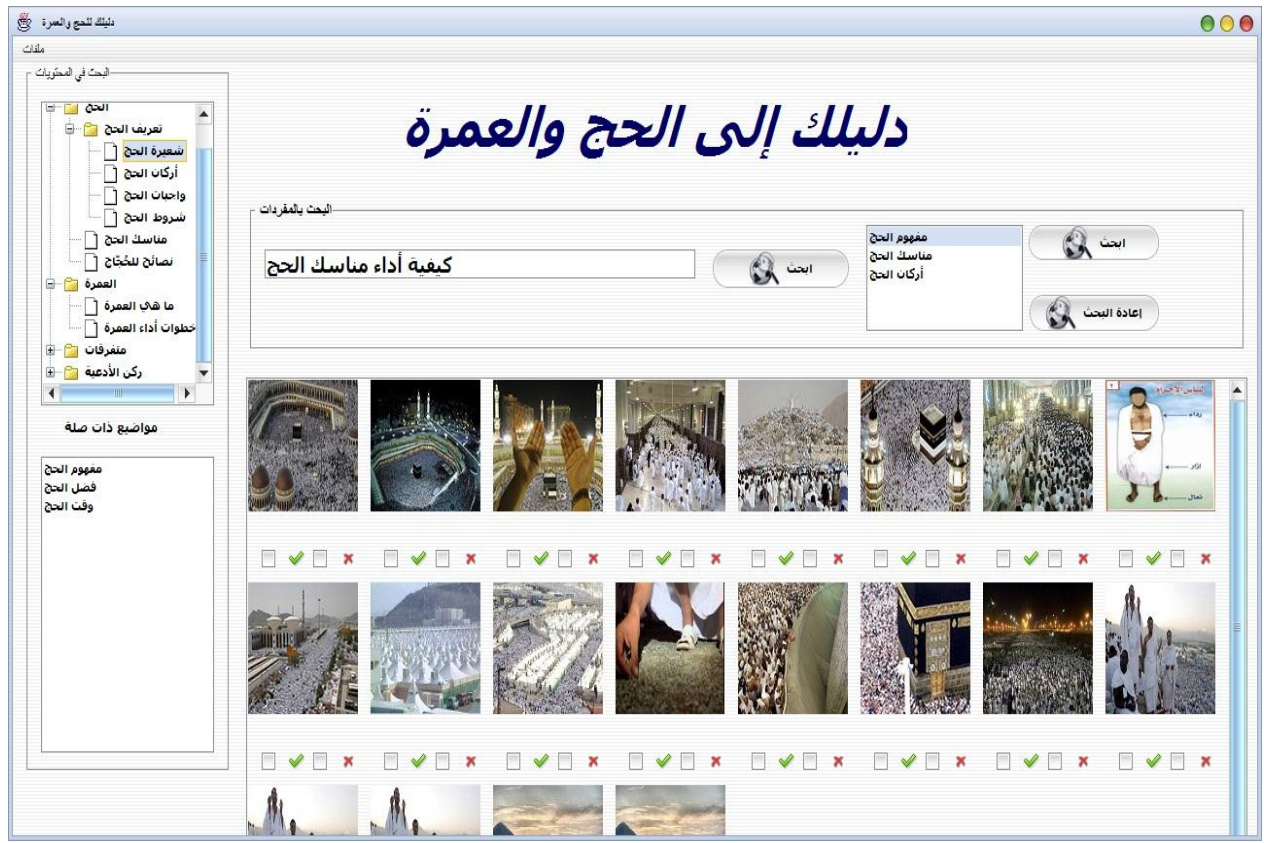

**Figure 5. Our multilingual Search Engine**

## 5. EXPERIMENTATION AND RESULTS

To properly evaluate our system, a database of 1000 videos illustrating rituals process is used. Each video is segmented to 2131 video key frames and two XML file. The structures of the XML files are described in Figure 6 and Figure 7. The first describes the 130 concepts with their corresponding plans (Figure 6). The second XML file describes the relationship context-concept (Figure 7):

```xml
<?xml version="1.0" encoding="UTF-8"?>
<concepts>
  <concept num="1" Name="عمرة">
  <concept num="2" Name="حج">
  <concept num="3" Name="جمرات">
  <concept num="4" Name="عرفة">
  <concept num="5" Name="طواف">
  <concept num="6" Name="السعي بين الصفا والمروة">
      <video Num="00001" Name="VIDEO_00001" Weight="0.91" NUMBER_shots="3" shotrepres="shot00001_1" />
      <video Num="00002" Name="VIDEO_00002" Weight="0.92" NUMBER_shots="6" shotrepres="shot00002_5" />
```

**Figure 6. XML file of concept description**

```xml
<?xml version="1.0" encoding="UTF-8" ?>
<contextes>
- <Contexte Num="1" Name="شعيرة الحج" Nbrconcept="3">
    <concept ConceptId="134" ConceptName="مفهوم الحج" Weight="1" />
    <concept ConceptId="135" ConceptName="فضل الحج" Weight="1" />
    <concept ConceptId="136" ConceptName="وقت الحج" Weight="1" />
  </Contexte>
```

**Figure 7. XML file of context description**

A user will launch an initial query $Q_0$ then do three iterations from this query. Among the first 30 videos returned by the system, the user will judge "m" relevant videos and "n" irrelevant videos. We use the recall which determines the possibility of the system to retrieve all relevant documents. It is the ratio between the number of retrieved documents and number of documents in the database for a given query. We also use the precision, which determines, for a given query, the system's ability to present only relevant documents. It is the ratio between the number of retrieved documents and number of documents returned (relevant and irrelevant).

By applying the precision and recall, we get the following curve:

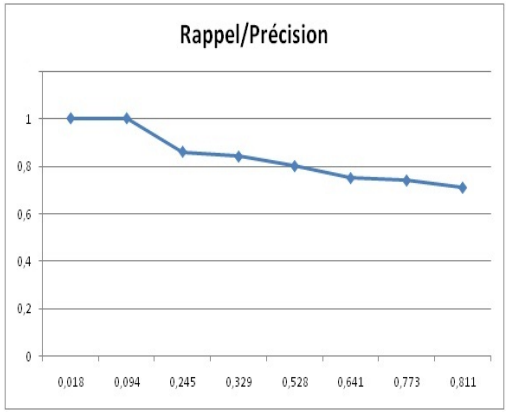

**Figure 10. Recall and precision corresponding to the original query "Q0"**

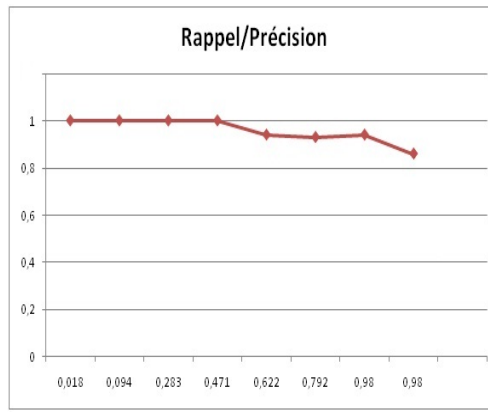

**Figure 9. Recall and precision corresponding to the query "Q1"**

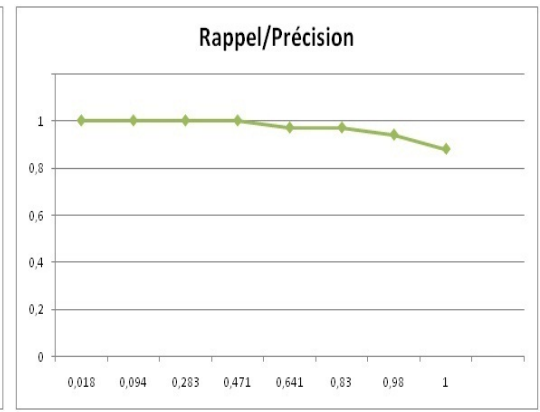

**Figure 8. Recall and precision corresponding to the query "Q2"**

According to Figure 10, we note that the accuracy begins to decrease from the point (0.094). This decrease is explained by the presence of irrelevant videos in the result from the point (0.094). According to precision values at different points, we can judge that the precision at the level of query "$Q_0$" is acceptable.

Now, we perform the same process to calculate the recall and precision for the new query "$Q_1$" from the initial query "$Q_0$" using the relevance feedback formula (8). We get the following result at Figure 8. Similarly, we will repeat the same process to calculate the recall and precision for a new query "$Q_2$" from previous query "$Q_1$". We get the following result at Figure 9.

$$PQ_i = P_{initial} + P\_fb_i \pm \alpha$$

Where : $PQ_0 = P_{initial} + P\_fb_0$ , $P\_fb_i = PQ_{i-1}$ , $P\_fb_0 = 0$, and $\alpha \pm 0,02$

According to Figure 9, we see stability in values clarification related to the first four points. This means that the relevant videos are located at the beginning of result. For the other points, the precision values are higher than 0.88 and later are superior to those of the query "$Q_1$".

## 6. Conclusion

In this paper, we present a new system for indexation and retrieval of audiovisual documents. Especially, we are interested by videos documents about rituals of Hajj and Umra. Our system propose a new approach based on exploiting ontology using several level. To help user in expressing the query, we integrate an expansion query process based on exploiting the same domain ontology. Our system is validated using 1000 videos. The validation proved that our system have an excellent precision and recall. We plan in the future to propose a mobile version of the system to facilitate it usage by pilgrims.

**References**

[1] M. Lew, N. Sebe, C Djerba, and R. Jain (2006), "Content-based Multimedia Information Retrieval: State of the Art and Challenges", ACM TOMCAPP vol.2, No. 1, pp. 1-19.

[2] M. Christel and A. Hauptmam : "The use and utility of high-level semantic features" CIVR, pages 134-144, 2005

[3] C. G. M. Snoek, M. Worring, D. C. Koelma, and A. W. M. Smeulders: ''A learned lexicon-driven paradigm for interactive video retrieval'', IEEE Transactions on Multimedia, pages 280-292, 2007.

[4] M. Worring, C. Snoek, O. de Rooji, G. P. Nguyen, R. Van Balen and D. Koelna: "Médiamill : Advanced browsing in news vidéo archives". CIVR, pages 533-536, 2006.

[5] V. Stefanos, M. Anastasia, K. Paul, D. Anastasios, M. Vasileios and K. Ioannis: "VERGE : A Video Interactive Retriveal Engine", 2010.

[6] Salton, G.; Buckley, C. (1988). "Term-weighting approaches in automatic text retrieval". Information Processing & Management. 24 (5): 513–523.

[7] M. Ben Halima, M. Hamroun, S. Moussa and A.M. Alimi, 2013 "An interactive engine for multilingual video browsing using semantic content", International Graphonomics Society Conference IGS, Nara Japan, pp 183-186.
.